\documentclass[sigconf]{acmart}

\usepackage{amsmath,amsfonts}
\usepackage{algorithm,algorithmic}
\usepackage{graphicx}
\usepackage{textcomp}
\usepackage{soul}
\usepackage{xcolor}

\setcopyright{none}

\begin{document}

\newcommand{\ph}[1]{{\textbf{#1:}}}
\newcommand{\mbf}[1]{\ensuremath{\mathbf{#1}}}

\title{SMILE: Robust Network Localization via Sparse and Low-Rank Matrix Decomposition}



\author{Lillian Clark}
\email{lilliamc@usc.edu}
\affiliation{%
    \institution{Electrical and Computer Engineering\\
    University of Southern California}
    \city{Los Angeles}
    \state{California}
    \country{USA}
}

\author{Sampad Mohanty}
\email{sbmohant@usc.edu}
\affiliation{%
    \institution{Computer Science\\
    University of Southern California}
    \city{Los Angeles}
    \state{California}
    \country{USA}
}

\author{Bhaskar Krishnamachari}
\email{bkrishna@usc.edu}
\affiliation{%
    \institution{Electrical and Computer Engineering\\
    University of Southern California}
    \city{Los Angeles}
    \state{California}
    \country{USA}
}


\begin{abstract}
Motivated by collaborative localization in robotic sensor networks, we consider the problem of large-scale network localization where location estimates are derived from inter-node radio signals. Well-established methods for network localization commonly assume that all radio links are line-of-sight and subject to Gaussian noise. However, the presence of obstacles which cause non-line-of-sight attenuation present distinct challenges.
To enable robust network localization, we present Sparse Matrix Inference and Linear Embedding (SMILE), a novel approach which draws on both the well-known Locally Linear Embedding (LLE) algorithm and recent advances in sparse plus low-rank matrix decomposition. We demonstrate that our approach is robust to noisy signal propagation, severe attenuation due to non-line-of-sight, and missing pairwise measurements. Our experiments include simulated large-scale networks, an 11-node sensor network, and an 18-node network of mobile robots and static anchor radios in a GPS-denied limestone mine. 
Our findings indicate that SMILE outperforms classical multidimensional scaling (MDS) which ignores the effect of non-line of sight (NLOS), as well as outperforming state-of-the-art robust network localization algorithms that do account for NLOS attenuation including a graph convolutional network-based approach.
We demonstrate that this improved accuracy is not at the cost of complexity, as SMILE sees reduced computation time for very large networks which is important for position estimation updates in a dynamic setting, e.g for mobile robots.  
\end{abstract}



\keywords{network localization, graph signal processing, low-rank matrix approximation}


\maketitle

\section{Introduction}
Robotic sensor networks operating in GPS-denied environments can benefit from collaborative localization ~\cite{6225016}.
When distance measurements between network nodes (e.g. mobile robots or static beacons) are available, the \textit{network localization} problem seeks to exactly recover the positions of each node in space. Intuitively, network localization algorithms leverage pairwise links to constrain the position estimate of every node in the network.
If the locations of some nodes are known, these nodes are considered \textit{anchors}. When sufficient conditions on the number of anchors and the graph induced by the distance measurements are met, we can recover the positions of all nodes ~\cite{goldenberg2005network}.

\begin{figure}
    \centering
    \includegraphics[width=0.8\columnwidth]{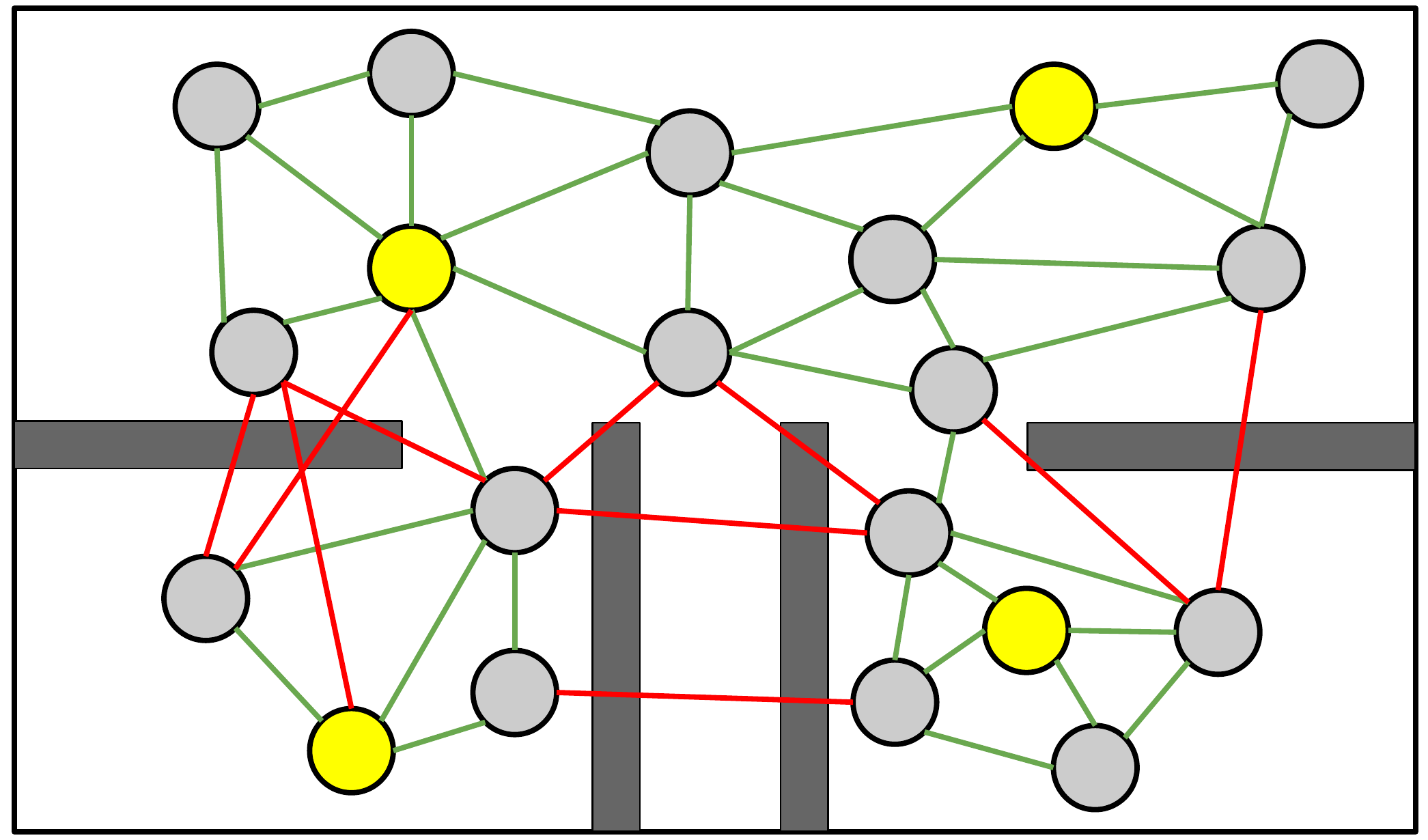}
    \caption{Illustration of network localization with NLOS. The objective is to accurately determine the localization of agents (grey) based on the known positions of anchors (yellow) given the constrains of noisy LOS links (green) and links with additional NLOS attenuation (red).}
    \label{fig:illustration}
\end{figure}


However, in realistic scenarios the distance measurements are typically noisy.
For example, radio signals can be used to estimate distance based on received signal strength, but signals are subject to noise from the wireless channel \citep{schwengler2016wireless}. 
When this noise is assumed to be Gaussian with zero-mean\footnote{Gaussian noise in dB is also referred to as log-normal fading.}, the redundancy offered by many links in a highly connected network serves to mitigate the problem. This explains why localization performance improves with the number of anchors used as reference points.

For many applications, including robotic exploration of unknown environments, signals see significant degradation from the presence of walls and obstacles. These non-line-of-sight (NLOS) links are difficult to model without an \textit{a priori} map but have a significant affect on the relationship between received signal strength (RSS) and distance \citep{fink2010online, clark2022prop}. Therefore, the ability to infer whether a pair of transmitters and receivers is NLOS can greatly improve network localization performance.

Additionally, when network nodes are spread over long distances, as may also be the case for robotic exploration, it is possible for the packets used to measure RSS to be dropped. When signal strength measurements are unavailable at long distances, our large-scale network localization approach must be robust to missing pairwise measurements.

\ph{Contribution}
In this work, we examine the network localization problem in the presence of noisy signals where pairwise measurements may be NLOS or missing.
We make two key observations: (1) the Euclidean distance matrix of the true node positions (in $d$ dimensions) will have rank $d+2$, which we show in Sec. \ref{sec:problem}, and (2) the matrix capturing the additional NLOS attenuation is commonly sparse and non-negative, i.e., there are a limited number of walls which can only degrade signal strength.
Given these observations, our approach draws on recent advances in sparse plus low-rank decomposition to first extract the positively biased NLOS attenuation, which we refer to as sparse matrix inference. After extracting the NLOS attenuation, we leverage a popular method for dimensionality reduction, locally linear embedding (LLE), to recover a set of weights which allow linear reconstruction to determine the exact positions of all nodes given the anchors' positions.

Specifically, we extend the discrete optimization approach for sparse matrix recovery which is presented in ~\cite{bertsimas2021sparse} to (1) handle missing pairwise measurements by imputation and (2) approximate the unknown sparsity structure of the sparse component of the matrix. Additionally, we provide details for solving for node positions using the LLE weight matrix directly (using least squares method) without the need for additional eigen-decomposition and coordinate frame alignment used conventionally. Our algorithm combines recent advances in sparse plus low rank decomposition and well established and interpretable algorithms like MDS and LLE and outperforms them in the face of NLOS attenuation.

\ph{Evaluation} 
We demonstrate that SMILE significantly improves localization accuracy over baseline methods which ignore the affect of NLOS attenuation.
Further, we draw parallels between this approach and another method for graph signal processing which has recently been demonstrated as promising, namely Graph Convolutional Networks (GCNs) ~\cite{yan2021graph}. We compare the performance of SMILE and a state-of-the-art GCN implementation on large-scale simulated networks and demonstrate an improvement in localization accuracy and reduced computation time for network of more than 1000 nodes.
Finally, we evaluate performance on two real-world networks: an outdoor wireless sensor network with 11 nodes, and 5 mobile robots and 13 static radios in a GPS-denied limestone mine with significant NLOS attenuation.

The paper is organized as follows. In Section \ref{sec:background} we provide a brief overview of seminal and recent work in network localization, and in Section \ref{sec:problem} we formally define the problem and notation.
SMILE is introduced and explained in detail in Section \ref{sec:approach}, including sparse matrix inference and subsequent steps for position estimation. We also draw parallels between our approach and the existing graph learning-based method.
In Section \ref{sec:results} we provide implementation details as well as experimental results on real and simulated data. We give concluding remarks in Section \ref{sec:conclusion}.

\section{Background and Related Work}
\label{sec:background}

\begin{table}
    \centering
    \caption{List of Abbreviations}
    \begin{tabular}{c|c}
        Abbr. & Description \\
        \hline
        EDM & Euclidean Distance Matrix \\
        LLE & Locally Linear Embedding\\
        MDS & Multidimensional Scaling \\
        NLOS & Non Line of Sight \\
        PSVD & Partial SVD\\
        RMSE & Root Mean Squared Error \\
        SDP & Semi Definite Programming \\
        SVD & Singular Value Decomposition
    \end{tabular}
    \label{tab:parameters}
\end{table}

\ph{Network localization}
Network localization is well-researched, and is commonly formulated as a least squares problem ~\cite{chowdhury2016advances,patwari2005locating,Beck2008}. Multidimensional scaling (MDS) and its extensions are popular methods for solving this problem ~\cite{saeed2018survey}. Classical MDS uses the distance matrix to compute a matrix of scalar products, typically called the Gram matrix, that captures pairwise correlation of the position vectors. The principal components from eigen-decomposition of this matrix are then used to recover relative node positions~\cite{bachrach2005localization}. Rather than compute over the entire distance matrix, Locally Linear Embedding (LLE) ~\cite{roweis2000nonlinear,saul2000introduction} applies principal component analysis to small neighborhoods, which improves performance when the reduction from the noisy (high-dimensional) data to the low-dimensional true positions is non-linear, and has shown promise in sensor network localization ~\cite{jain2015locally}.

\ph{Exploiting sparsity} 
If the data is well-described by a particular statistical model (e.g. Gaussian or log-normal), we can instead form the maximum likelihood estimation problem ~\cite{patwari2003relative}, and solve using semi-definite programming (SDP) methods ~\cite{Biswas2006approach, biswas2004semidefinite}. 
Recent works extend SDP methods to consider non-Gaussian noise ~\cite{yin2015cooperative} and, more specifically, NLOS noise ~\cite{marano2010nlos,chen2011non,jin2020exploiting}. Jin \textit{et al.} demonstrate that when the NLOS noise has a certain structure, namely non-negative and sparse, a sparsity-promoting term in the objective function can improve the performance of SDP approaches ~\cite{jin2020exploiting}. However, SDP methods suffer with respect to complexity and are intractable for very large networks ~\cite{yan2021graph}.

\ph{Sparse and low rank decomposition}
Another approach is to recover the matrix of NLOS attenuation directly.
In recent years, sparse and low-rank matrix recovery has drawn attention due to its relevance in signal processing, statistics, and machine learning ~\cite{wen2018survey}. Bertsimas \textit{et al.} recently proposed a discrete optimization approach to sparse and low-rank recovery which uses alternating minimization~\cite{bertsimas2021sparse}. Our work leverages this method, extends it to the case of missing measurements and unknown sparsity, and demonstrates that it can serve as an important component of a robust network localization algorithm.

\ph{Graph convolutional networks}
To effectively exploit the relational information of graph-structured data, graph neural networks (GNNs) have recently become a popular method for approaching optimization problems in wireless networks~\cite{mao2019learning}. Yan \textit{et al.} recently demonstrated promising results in the application of Graph Convolutional Networks (GCNs) to the network localization problem. Their approach maintains accurate localization despite NLOS attenuation, and is scalable to large-scale networks at an affordable computation cost. However, we will see that the learned model is unable to exactly recover positions for a completely observed distance matrix in the absence of noise. In this work we propose a novel network localization algorithm which builds on the principles of multidimensional scaling for exact recovery in the absence of noise, exploits the sparsity of NLOS attenuation for improved localization accuracy, and scales to very large networks at an affordable computation cost.

\section{Problem Formulation}
\label{sec:problem}
Let $A$ be the set of anchors whose positions are known, where $|A| = n_A$. Let $B$ be the set of agents whose positions are unknown, where $|B| = n_B$. Let $C = A \cup B$ be the set of nodes which includes both anchors and agents, with $n_A + n_B = n$.
We define the $(n \times n)$ true distance matrix $\mathbf{D}$ as 
$$
    \mathbf{D}[i,j] = || \mathbf{p}_i - \mathbf{p}_j ||
$$
where $\mathbf{p}_i = [x_i, y_i]^T$ is the position of node $i$. We establish the convention that the first $n_A$ rows and $n_A$ columns pertain to the anchors. $\mathbf{D}^{\circ 2}$ is the Euclidean distance matrix (EDM, containing squared distances i.e $\mbf{D}$ squared entry-wise), which is zero along the diagonal and symmetric~\cite{gower1985properties}. If matrix $\mathbf{P} \in \mathbb{R}^{n \times 2}$ with $\mathbf{P}[i,:] = \mathbf{p}_i^T$ be the position matrix, we can see that $\mbf{D^{\circ 2}}$ has rank $4$. In general, the EDM of a configuration of points embedded in  $\mathbb{R}^d$ has rank at most $d+2$ (and exactly $d+2$ if the points are in general position as opposed to more special or coincidental cases, e.g., points in 3D which lie on a line). This is briefly justified below, with further details provided in \cite{dokmanic2015euclidean}.
\begin{align*}
    \mathbf{D}[i,j]^2 = ||\mathbf{p}_i - \mathbf{p}_j||^2 \\
    =  \langle \mathbf{p}_i - \mathbf{p}_j , \mathbf{p}_i - \mathbf{p}_j \rangle \\ 
    = \langle \mathbf{p}_i, \mathbf{p}_i \rangle + \langle \mathbf{p}_j , \mathbf{p}_j \rangle - 2 \langle \mathbf{p}_i , \mathbf{p}_j \rangle \\
    = \mathbf{p}_i^T\mathbf{p}_i + \mathbf{p}_j^T\mathbf{p}_j - 2 \mathbf{p}_i^T \mathbf{p}_j\\
    \implies \mathbf{D^{\circ 2}} = \underbrace{\mathbf{diag(PP^T) 1^T}}_{rank-1} + \underbrace{\mathbf{1 diag(PP^T)}^T}_{rank-1} - \underbrace{2\mathbf{PP^T}}_{rank-2}
\end{align*}
Since $rank(\mathbf{A} + \mathbf{B}) \leq rank(\mathbf{A}) + rank(\mathbf{B})$ and $rank(\mathbf{CC^T}) = rank(\mbf{C^TC}) = rank(\mathbf{C})$, we have $\mathbf{D}^{\circ 2}$ as low-rank with rank = 4. 

Let $\mathbf{O}$ be the matrix that contains the distances that we observe, where $\mathbf{O}[i,j]$ is a function of the strength of radio signal transmitted by node $i$ and received by node $j$. In general, our observation can be captured as
$$
    \mathbf{O} = \mathbf{\Omega} \circ [\mathbf{D} + \mathbf{N} + \mathbf{S} ] 
$$
where $\mathbf{\Omega}$ is the observation mask and takes on values of 1 when a measurement is available and 0 otherwise (for instance, when the transmitter and receiver are out of range). $\mathbf{N}$ is an asymmetric matrix capturing noise in the observations, which we assume is Gaussian and zero-mean. $\mathbf{S}$ captures the additional NLOS attenuation, and as in ~\cite{jin2020exploiting} we assume $\mathbf{S}$ is non-negative and sparse. We also assume $\mathbf{S}$ is symmetric which relies on assumptions that an attenuating obstacle will affect a radio link in both directions equally.

We make the realistic assumption that nodes are in general position, meaning that in 2D they do not lie on a straight line. We also assume that an upper bound on the distance between any two nodes, $d_\textrm{max}$, is known or can be approximated.
Our objective is finding an estimate for the locations of all nodes $\mathbf{\hat{P}} = [\mathbf{\hat{p}}_1, ... ,\mathbf{\hat{p}}_n]^T$ which is consistent with the prior information - (1) the observations in $\mbf{O}$ as well as (2) the known anchor positions $\mathbf{P_A} = [\mathbf{p}_1, ... ,\mathbf{p}_{n_A}]^T$. 


\section{SMILE}
\label{sec:approach}

\begin{figure}
    \centering
    \includegraphics[width=\columnwidth]{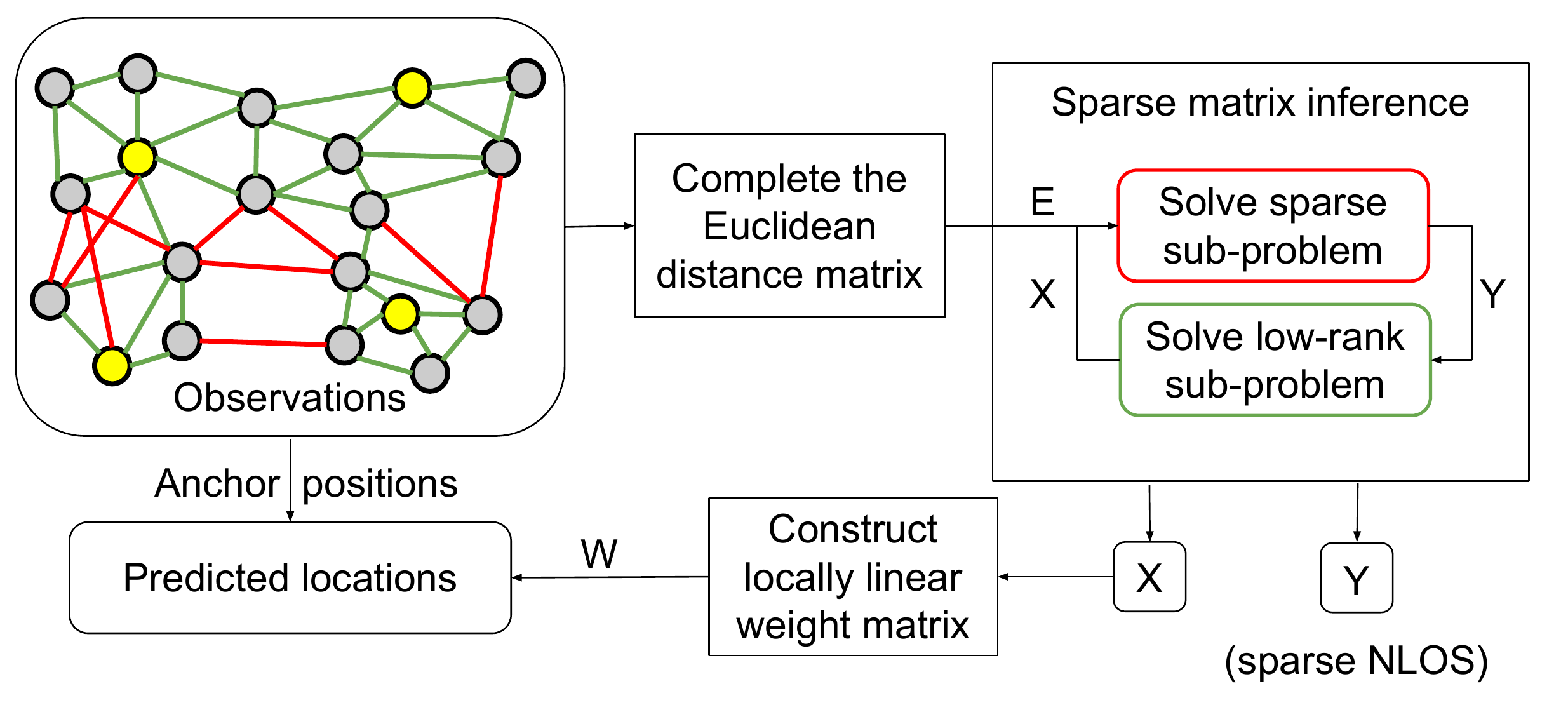}
    \caption{An overview of Sparse Matrix Inference and Linear Embedding (SMILE).}
    \label{fig:smile_block_diagram}
\end{figure}

In this section we propose Sparse Matrix Inference and Linear Embedding, our novel large-scale network localization algorithm, which is illustrated in Fig. \ref{fig:smile_block_diagram}. Details of the algorithm are presented in Algorithm \ref{alg:approach}. In subsection \ref{sec:smi}, we discuss our method of extracting NLOS attenuation via sparse matrix inference, which results in a low-rank approximation of the Euclidean distance matrix. Then in subsection \ref{sec:le}, we discuss our method of transforming this matrix into an estimate of the locations of all nodes.

\subsection{Sparse Matrix Inference}
\label{sec:smi}

For a given input matrix $\mathbf{E} \in \mathbb{R}^{n \times n}$, for which $\mathbf{E} = \mathbf{X} + \mathbf{Y}$, sparse matrix inference seeks to find the low-rank component $\mathbf{X} \in \mathbb{R}^{n \times n}$ and sparse component $\mathbf{Y} \in \mathbb{R}^{n \times n}$ which solves:
\begin{align}
    \begin{split}
    \label{eq:f}
    \min_{\mathbf{X}, \mathbf{Y}} f(\mathbf{X}, \mathbf{Y}) = ||\mathbf{E} - \textbf{X} - \textbf{Y}||_F^2 + \lambda||\textbf{X}||_F^2 + \mu||\textbf{Y}||_F^2
    \\
    \textrm{s.t.} \quad \textrm{rank}(\mathbf{X}) \leq \alpha, ||\mathbf{Y}||_0 \leq \beta
    \end{split}
\end{align}
where $||\mathbf{X}||_F$ denotes the Frobenius norm. In this subsection we describe how to decompose $\mathbf{E}$ $\mathbf{X}$ and $\mathbf{Y}$ using alternating minimization, and how to use this technique on the problem defined in Sec. \ref{sec:problem}.

Firstly, we square the observed matrix (line 2). Temporarily assuming $\mathbf{N}=\mathbf{0}$ and no missing observations, i.e $\mbf{\Omega} = \mbf{11^T}$,
\begin{align}
\label{NoNoiseForm}
\begin{split}
    \mathbf{E} = \mathbf{O}^{\circ 2} = (\mbf{D} + \mbf{S})^{\circ 2} \\ 
    =  \underbrace{\mathbf{D}^{\circ 2}}_{\text{low-rank}} + \underbrace{\mathbf{S}^{\circ 2} + 2\mathbf{D} \circ \mathbf{S}}_{\text{sparse}} \\ 
    \end{split}
\end{align}
where the expression in parentheses is non-negative and sparse (because $\mathbf{S}$ is non-negative and sparse), and $\mathbf{D}^{\circ 2}$ is low-rank. Thus the problem is amenable to the sparse matrix inference framework. 

Let us consider the case with noise $N \neq 0$ and no missing observations
\begin{align}
    \begin{split}
        \mbf{E} = \mbf{O}^{\circ 2} = (\overbrace{\mbf{D + N}}^{\mbf{\tilde{D}}} + \mbf{S})^{\circ 2} = (\mbf{\tilde{D}} + \mbf{S})^{\circ 2} \\
        = \mbf{\tilde{D}^{\circ 2}} + \underbrace{\mbf{S^{\circ 2} + \tilde{D} \circ S}}_{\text{sparse}}
    \end{split}
\end{align}
The last expression is exactly the same as the formulation in \ref{NoNoiseForm} except that $\mathbf{\tilde{D}}^{\circ 2} = \mbf{D^{\circ 2} + N^{\circ 2} + 2 N \circ D} $ may no longer be low rank due to the addition of $\mbf{N^{\circ 2} + 2 N \circ D}$ where entries of the first term are now i.i.d from a Chi-Squared Distribution $\mbf{\mathcal{X}_1^2}$. However, in our ablation study, we empirically verify that as long as the standard deviation of the normal noise in entries of $\mbf{N}$ is not too large, using the estimated low-rank matrix from the sparse matrix plus low-rank inference, SMILE is able to recover $\mbf{P}$ fairly faithfully under RMSE loss.

\ph{Missing measurements} The approach presented in ~\cite{bertsimas2021sparse} assumes a complete input matrix. While more complex approaches to matrix completion exist, for example by finding the sum total length of the shortest path between two nodes ~\cite{shang2004localization,shang2004improved}, we take a simpler approach. When $\mbf{\Omega} \neq \mbf{11^T}$, we complete the observation matrix by filling in missing values with $d_\textrm{max}$ (line 1). Even in the face of missing observations, we show that the problem remains amenable to the sparse matrix inference framework as long as the missing observations are not too large of a fraction of the total number of observations in $\mbf{O}$. 
\begin{align}
    \begin{split}
        \mbf{E} = \mbf{\tilde{O}}^{\circ 2} = [\mbf{\Omega \circ O}]^{\circ 2}\\
        = [\mbf{\Omega} \circ (\mbf{D + N + S}) + (\mbf{1-\Omega})d_{max}]^{\circ 2}  \\
        = [\mbf{\Omega \circ D + \Omega \circ (N+S) + (1- \Omega)}d_{max}]^{\circ 2} \\
        = [\mbf{D - (1- \Omega) \circ D + \Omega \circ (N + S) + (1- \Omega)}d_{max}]^{\circ 2} \\
        = [\mbf{D + \Omega \circ N} + \underbrace{\mbf{\Omega \circ S  + (1-\Omega)}\circ (d_{max}\mbf{I - D})}_{\tilde{S}\text{ (sparse)}}]^{\circ 2} \\
        = [\underbrace{\mbf{D + \Omega \circ N}}_{\mbf{\tilde{D}}} + \mbf{\tilde{S}}]^{\circ 2} 
        = [\mbf{\tilde{D} + \tilde{S}}]^{\circ 2} \\
        = \mbf{\tilde{D}^{\circ 2}} + \underbrace{\mbf{\tilde{S}^{\circ 2} + 2\tilde{D}\circ\tilde{S}}}_{\text{sparse}}
    \end{split}
\end{align}
Thus, $\mathbf{\tilde{O}}$ decomposes similarly to $\mathbf{O}$ as long as $(1-\mathbf{\Omega}) + \mathbf{S}$ is still sparse, i.e. the total number of NLOS and missing measurements is low. $\mbf{\tilde{D}}^{\circ 2}$ is still amenable to SMILE like in the previous case.

\ph{Alternating minimization}
As presented in ~\cite{bertsimas2021sparse}, we alternate between solving two sub-problems. 
For a given $\mathbf{X}$, we estimate $\mathbf{Y}$ by composing a sparse matrix with non-zero entries at the largest indices of $(\mathbf{E} - \mathbf{X})$. This is described further in Algorithm \ref{alg:compose_sparse}.
Then for a given $\mathbf{Y}$, we estimate $\mathbf{X}$ by reducing the rank of $(\mathbf{E} - \mathbf{Y})$. 
This repeats until the value of the objective function in Eq. \ref{eq:f} has converged, corresponding to the inner loop (lines 11-18). We initialize $\mathbf{X}$ with a random, low-rank matrix and repeat this process for $T > 0$ random initialization, ultimately selecting the decomposition which minimizes the objective function (lines 5-19).

\renewcommand{\algorithmicrequire}{\textbf{Input:}}
\renewcommand{\algorithmicensure}{\textbf{Output:}}

\begin{algorithm}
    \caption{SMILE}
    \label{alg:approach}
    \begin{algorithmic}[1]
        \REQUIRE $\mathbf{O}$ (observation matrix), $\mathbf{P}_A$ (anchor positions), $\alpha$ (desired rank)
        \ENSURE $\mathbf{\hat{P}_B}$ (agent location estimates)
        \STATE $\mathbf{\tilde{O}} = \mathbf{O} + (1+\mathbf{\Omega}) d_\textrm{max}$
        \STATE $\mathbf{E} = \tilde{\mathbf{O}}^{\circ 2}$
        \STATE $\hat{\beta} \leftarrow$ $\hat{\beta}_i$
        \WHILE {$\Delta f/f > \epsilon_{\beta}$}
            \FOR {$t$ in $\{1, ... , T\}$}
                \STATE $\mathbf{X}' \in \mathbb{R}^{n \times n} \leftarrow$ random
                \STATE $\mathbf{U}, \mathbf{\Sigma}, \mathbf{V}^T = \textrm{PSVD}(\mathbf{X}', \alpha)$
                \STATE $\mathbf{X} = \mathbf{U} \mathbf{\Sigma} \mathbf{V}^T$
                \STATE $\mathbf{Y} \in \mathbb{R}^{n \times n} \leftarrow \mathbf{0}$
                \STATE $f = ||\mathbf{E} - \textbf{X} - \textbf{Y}||_F^2 + \lambda||\textbf{X}||_F^2 + \mu||\textbf{Y}||_F^2$
                \WHILE {$\Delta f/f > \epsilon$}
                    \STATE $\mathbf{Y}' =$ compose\_sparse($\mathbf{E} - \mathbf{X}, \hat{\beta}$)
                    \STATE $\mathbf{Y} = \frac{1}{1+\mu} \mathbf{Y}' \circ (\mathbf{E} - \mathbf{X})$
                    \STATE $\mathbf{X}' = \frac{1}{1+\lambda} (\mathbf{E} - \mathbf{Y})$
                    \STATE $\mathbf{U}, \mathbf{\Sigma}, \mathbf{V}^T = \textrm{PSVD}(\mathbf{X}', \alpha)$
                    \STATE $\mathbf{X} = \mathbf{U} \mathbf{\Sigma} \mathbf{V}^T$
                    \STATE $f = ||\mathbf{E} - \textbf{X} - \textbf{Y}||_F^2 + \lambda||\textbf{X}||_F^2 + \mu||\textbf{Y}||_F^2$
                \ENDWHILE
            \ENDFOR
            \STATE $\hat{\beta} \leftarrow \hat{\beta} + \eta$
        \ENDWHILE
        \STATE $\mathbf{W} \in \mathbb{R}^{n \times n} \leftarrow \mathbf{0}$
        \FOR {each node $i$}
            \STATE find $k$ nearest neighbors $NN(i)$
            \FOR {each pair of neighbors ($j$, $l$)}
                \STATE $\mathbf{H} \in \mathbb{R}^{k \times k}$ where $\mathbf{H}[j,l] = \frac{1}{2} (\mathbf{X}[i,l] + \mathbf{X}[j,i] - \mathbf{X}[j,l])$
                \STATE solve $\mathbf{H} \mathbf{w} = 1$ for $\mathbf{w}$
            \ENDFOR
            \STATE $\mathbf{W}_i \leftarrow \mathbf{w}/\sum{\mathbf{w}}$ at indices of neighbors
        \ENDFOR 
        \STATE $\mathbf{m} = (\mathbf{I}-\mathbf{W})_A \mathbf{P}_A$
        \STATE solve $(\mathbf{W}-\mathbf{I})_B \mathbf{\hat{P}}_B = \mathbf{m}$ for $\mathbf{\hat{P}}_B$
        \RETURN $\mathbf{\hat{P}_B}$
    \end{algorithmic}
\end{algorithm}

\begin{algorithm}
    \caption{compose\_sparse}
    \label{alg:compose_sparse}
        \begin{algorithmic}[1]
          \REQUIRE $\mathbf{M} \in \mathbb{R}^{n \times n}$ (matrix), $\beta$ (sparsity)
          \ENSURE $\mathbf{B}$ (binary matrix)
          \STATE sorted indices = argsort($\mathbf{M}$)
          \STATE one indices = sorted indices[-$\beta$:]
          \STATE $\mathbf{B} \in \mathbb{R}^{n \times n} \leftarrow \mathbf{0}$ 
          \STATE $\mathbf{B}$[one indices] = 1
          \RETURN $\mathbf{B}$
        \end{algorithmic}
\end{algorithm}

\begin{table}
    \centering
    \caption{SMILE parameters}
    \begin{tabular}{c|c}
        Parameter & Symbol \\
        \hline
        $k$ & Number of neighbors \\
        $\hat{\beta}_i$ & Initial sparsity estimate \\
        $\eta$ & Step size \\
        $T$ & Number of random initializations \\
        $\lambda$ & Low-rank matrix regularizer \\
        $\mu$ & Sparse matrix regularlizer \\
        $\epsilon$ & Inner loop tolerance \\
        $\epsilon_{\beta}$ & Outer loop tolerance \\
    \end{tabular}
    \label{tab:parameters}
\end{table}

\begin{figure}
    \centering
    \includegraphics[width=\columnwidth]{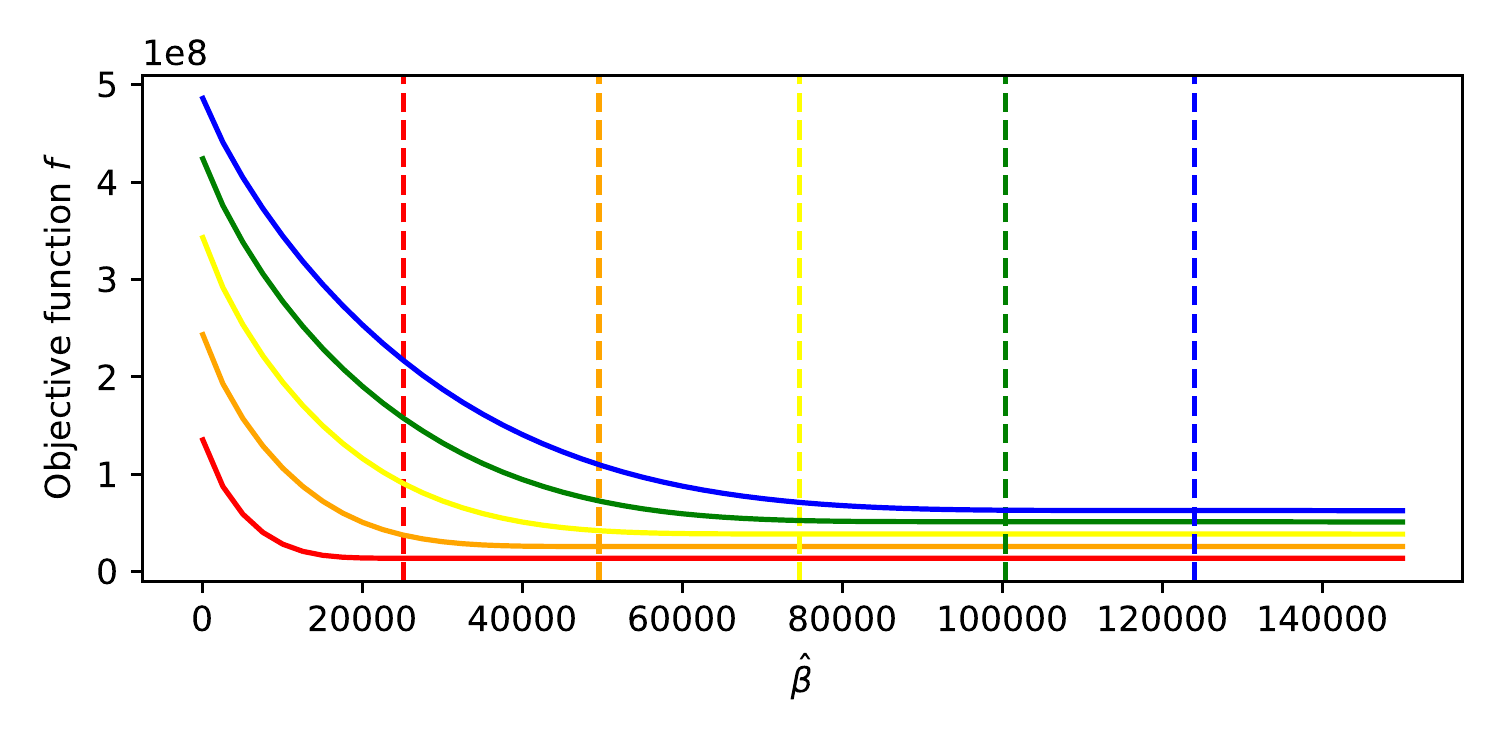}
    \caption{The final value of the sparse matrix inference objective function $f$ (Eq. \ref{eq:f}) when we assume different levels of sparsity $\hat{\beta}$, plotted alongside the true sparsity $\beta$ (dotted line) for various simulated matrices $\mathbf{E} \in \mathbb{R}^{500 \times 500}$.}
    \label{fig:k1_and_convergence}
\end{figure}

\ph{Unknown sparsity} The approach presented in ~\cite{bertsimas2021sparse} assumes the sparsity of $\mathbf{Y}$ is known, however in realistic settings this may not be true. We observed empirically that for a matrix with known components, solving the sparse matrix inference problem for increasing estimate $\hat{\beta}$ causes the objective function to decrease, and around the true value of $\beta$ it converges (Fig. \ref{fig:k1_and_convergence}. This motivates the outer loop (lines 3-21). 
To reduce computation time, we can initialize $\hat{\beta}$ with the approximate sparsity, increase the search step size $\eta$, or increase the converge tolerance $\epsilon_\beta$. However this may come at the cost of localization accuracy.


\subsection{Linear Embedding}
\label{sec:le}

Next we find the set of k nearest neighbors $NN(i)$ for each node $i$ using the estimated  EDM gotten from sparse matrix inference.
For each node, we compute a $k \times k$ matrix $\mathbf{H}$ which captures the  pairwise similarity between the nodes in $NN(i) \cup \{i\}\;$(line 25) using the corresponding sub-matrix in the estimated EDM. $\mathbf{H}$ is calculated using the same approach used for classical MDS, but for local neighborhoods ~\cite{bachrach2005localization}. 
From $\mathbf{H}$ we can recover the desired weight matrix $\mathbf{W}$ (lines 26-27), first by solving for $\mathbf{w}$ and then inserting it at the row position corresponding to the node index of $i$ with the entries in $\mathbf{w}$ aligned with the column belonging to the nearest neighbours and zeros for non-neighbours.
At this point, each row of $\mathbf{W}$ corresponds to a node and captures how the node's position can be expressed as a linear combination of its neighbors. This is a meaningful representation of the positions of all nodes, but is not yet a node embedding. Typically, LLE then computes a sparse matrix $(\mathbf{I}-\mathbf{W})^T (\mathbf{I}-\mathbf{W})$ whose eigenvectors corresponding to the two smallest non-zero eigenvalues result in a solution up to rotation and translation.

\ph{Using the anchors' positions}
In this setting, we can compute a solution using $\mathbf{W}$ and the anchor location $\mathbf{P}_A$ directly. This is possible because, by construction, $\mathbf{W} \mathbf{P} = \mathbf{P}$. This is essentially a system of $n$ equations with $n_B < n$ unknowns. Treating $\mbf{W - I} \in \mathbb{R}^{n \times n}$ as a block matrix, let $(\mathbf{I}-\mathbf{W})_A = \mbf{(I - W)[:,:n_A]} \in \mathbb{R}^{n \times n_A}$ be the sub-matrix corresponding to the anchors and $(\mathbf{W}-\mathbf{I})_B = \mbf{(I - W)[:,-n_B:]} \in \mathbb{R}^{n \times n_B}$ be the sub-matrix corresponding to the agents, where $\mathbf{I} \in \mathbb{R}^{n \times n}$ is the identity matrix. Similarly treating $P \in \mathbb{R}^{n \times 2}$ as a block matrix, let $P_A = P[:n_A,:] \in \mathbb{R}^{n_A \times 2}$ and $P_B = P[-n_B,:] \in \mathbb{R}^{n_B \times 2}$ correspond to the block matrices for positions of anchors and agents respectively. 
With some manipulation we get,
\begin{align}
\begin{split}
    \mbf{WP = P}
   \implies \overbrace{\mbf{(W-I)}}^{T}\mbf{P = 0}  \implies \mbf{TP=0}\\
   \implies \begin{bmatrix} \mbf{T_A} & | & \mbf{T_B} \end{bmatrix} \begin{bmatrix} \mbf{P_A} \\ -\\ \mbf{P_B} \end{bmatrix} = \mbf{0} \\
   \implies \mbf{T_A P_A + T_B P_B = 0}
   \implies   \mbf{T_AP_A = -T_BP_B} \\
    \implies \underbrace{(\mathbf{I}-\mathbf{W})_A \mathbf{P}_A}_{\mbf{m}\text{, known}} = \underbrace{(\mathbf{W}-\mathbf{I})_B}_{\text{known}} \underbrace{\mathbf{P}_B}_{\text{unknown}}    
\end{split}
\end{align}
(lines 29-30) and from the last expression we can estimate a solution $\mbf{\hat{P}_B}$ for the unknown agent positions using the least squares minimization.

\subsection{Comparison with Graph Convolutional Networks}
We highlight several interesting parallels between SMILE and GCN. 
Briefly, the approach presented by Yan \textit{et al.} computes the position estimates
\begin{equation}
    \mathbf{\hat{P}} = \mathbf{A'} \phi ( \mathbf{A'} ( \mathbf{A} \odot \mathbf{O}) \mathbf{Z}^{(1)} ) \mathbf{Z}^{(2)}
\end{equation}
where $\mathbf{A}$ is the thresholded adjacency matrix for a given threshold $\theta_{GCN}$, $\mathbf{A'}$ is the row-normalized augmented thresholded adjacency matrix, $\phi(\cdot)$ is a nonlinear activation function, and $\mathbf{Z}^{(1)}, \mathbf{Z}^{(2)}$ are learned weights.
Note that the graph signal is the observed matrix. 

\begin{enumerate}
    \item \ph{$\theta_{GCN}$ and $k$}
    This approach introduces a threshold such that edges between nodes in the graph are present only if the observed distance is less than the threshold. Decreasing this threshold is similar to decreasing the number of nearest neighbors, and both have the benefit of noise truncation.
    However, as the threshold is increased, GCN experiences over-smoothing due to the aggregation step. Intuitively, if every node is connected to every other node, the aggregation step causes all node embedding to collapse at a single point. This means that more information, i.e., additional pairwise measurements, actually hurts the algorithm.
    SMILE does not experience this issue, and we will see in Fig. \ref{fig:vary_n_neighbors} that increasing $k$ improves performance at the cost of increased runtime.
    \item \ph{Low-pass filtering and PSVD}
    Repeated multiplication by the normalized adjacency matrix acts as a low pass filter ~\cite{nt2019revisiting}. This reduces the rank of the graph signal, similar to the process of extracting the low-rank component of the noisy euclidean distance matrix. However, repeated multiplication by the normalized adjacency matrix reduces the rank of the observed matrix by some amount, which corresponds to distances (rather than squared distances). Our approach applies rank reduction to exactly reduce the Euclidean distance matrix to rank $d+2$, informed by the principles of the problem formulation.
    \item \ph{Convolution and linear embedding}
    Each convolution layer multiplies the adjacency matrix and input matrix by a set of weights. Intuitively, this makes the node features at the next layer a weighted sum of the neighbor features. The learned GCN weights are analogous to the set of weights $\mathbf{W}$ in SMILE which allow linear reconstruction of agents from the anchors positions. However, the repeated convolutions and nonlinear activation prevent a straightforward analysis. The interpretability of $\mathbf{W}$ is an advantage of our approach.
\end{enumerate}

Qualitatively, we expect SMILE to outperform GCN because (1) it is not subject to oversmoothing and makes careful and productive use of additional pairwise measurements, (2) it finds a Euclidean distance matrix with the exact expected rank, and (3) it relies on principled methods to determine a weight matrix relating nodes to their neighbors, which is thus interpretable.
In the next section, we compare these approaches quantitatively.

\section{Results}
\label{sec:results}
In this section we evaluate the performance of SMILE with respect to localization accuracy and computation time. We also consider performance under different noise settings, as an ideal method is both robust to high-levels noise and able to exactly recover positions when possible. Localization accuracy is measured by the root-mean-squared-error (RMSE) given by $||\mathbf{P}_B - \mathbf{\hat{P}}_B ||_F$. Robustness considers the affect of Guassian noise, NLOS attenuation, and missing pairwise measurements.
The experiments consider networks in 2D, thus the desired rank of EDM is $\alpha=4$. 
The SMILE parameters are set to $k=50, \hat{\beta}_i=5 \frac{n^2}{100}, \eta=\frac{n^2}{100}, T=1, \lambda=0.01, \mu=0.1, \epsilon=0.001$, and $\epsilon_\beta=0.01$. 
For comparison, we train a two-layer GCN according to ~\cite{yan2021graph} with distance threshold 1.2.
More details of our implementation are available online \footnote{https://github.com/ANRGUSC/smile-network-localization}.

\subsection{Simulation results}
Our simulated scenarios consider $n$ nodes randomly placed over a 5m $\times$ 5m square area, with the first $n_A$ nodes considered anchors. Noise $\mathbf{N}$ is drawn from a zero-mean Gaussian, $\mathbf{N}[i,j] \sim \mathcal{N}(0, \sigma^2)$. 
NLOS noise is drawn from a uniform distribution, $\mathbf{S}[i,j] = \mathbf{S}[j,i] \sim \mathcal{U}[0,10]$ with probability $p_\textrm{NLOS}$ and 0 otherwise. 
Both matrices are zero along the diagonal, $\mathbf{N}[i,i] = \mathbf{S}[i,i] = 0$.
The observation mask limits our observed distance measurements by a threshold $\theta$ such that $\mathbf{\Omega}[i,j] = 1$ if $\mathbf{O}[i,j] \leq \theta$ and 0 otherwise.

\begin{figure}
    \centering
    \includegraphics[width=\columnwidth]{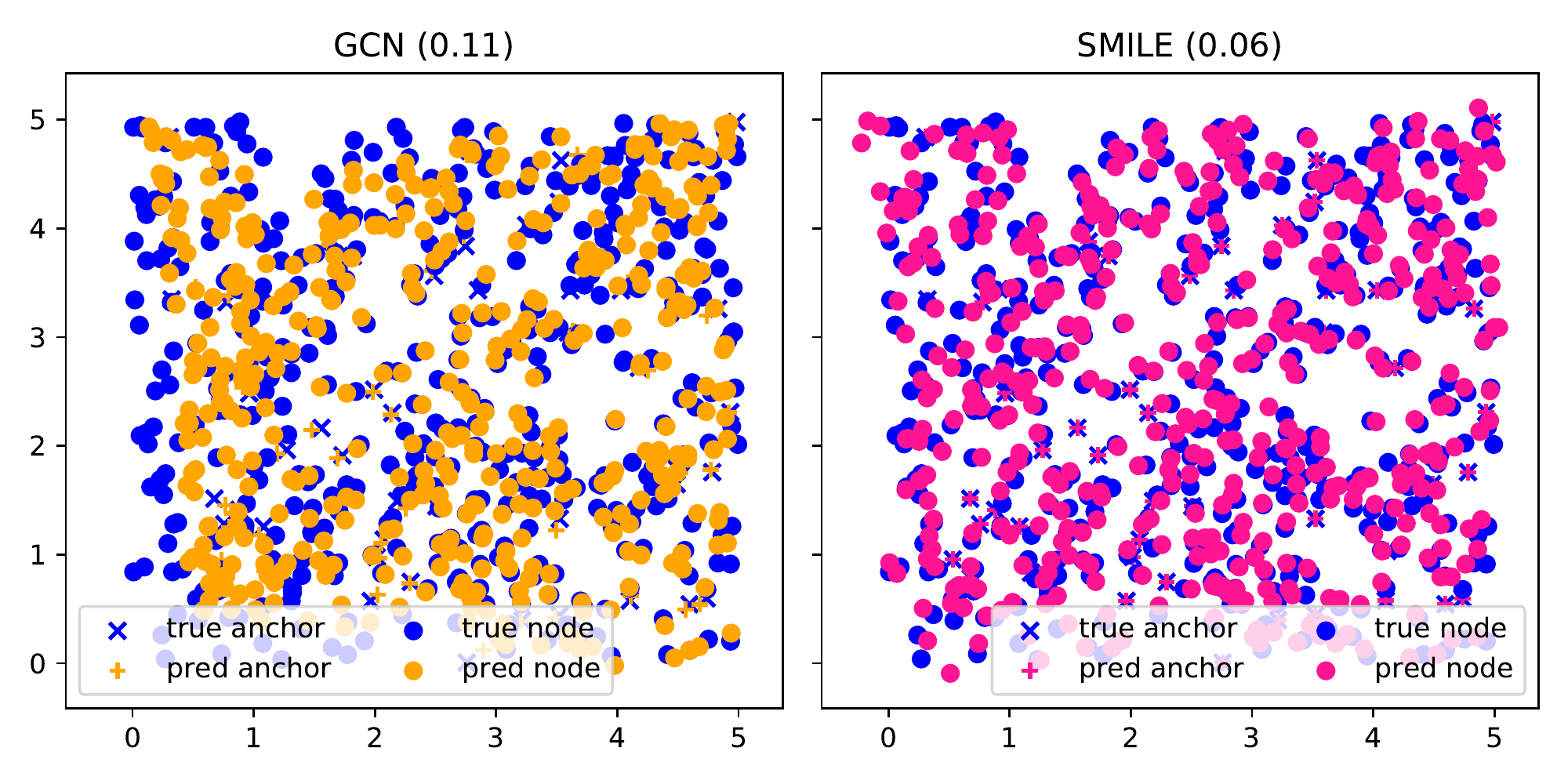}
    \caption{Localization with GCN (Yan et al.~\cite{yan2021graph}) and novel SMILE for a 500 node network with 50 anchors. GCN achieves RMSE 0.11 and SMILE achieves RMSE 0.06.}
    \label{fig:compare_with_gcn}
\end{figure}

\begin{figure}
    \centering
    \includegraphics[width=\columnwidth]{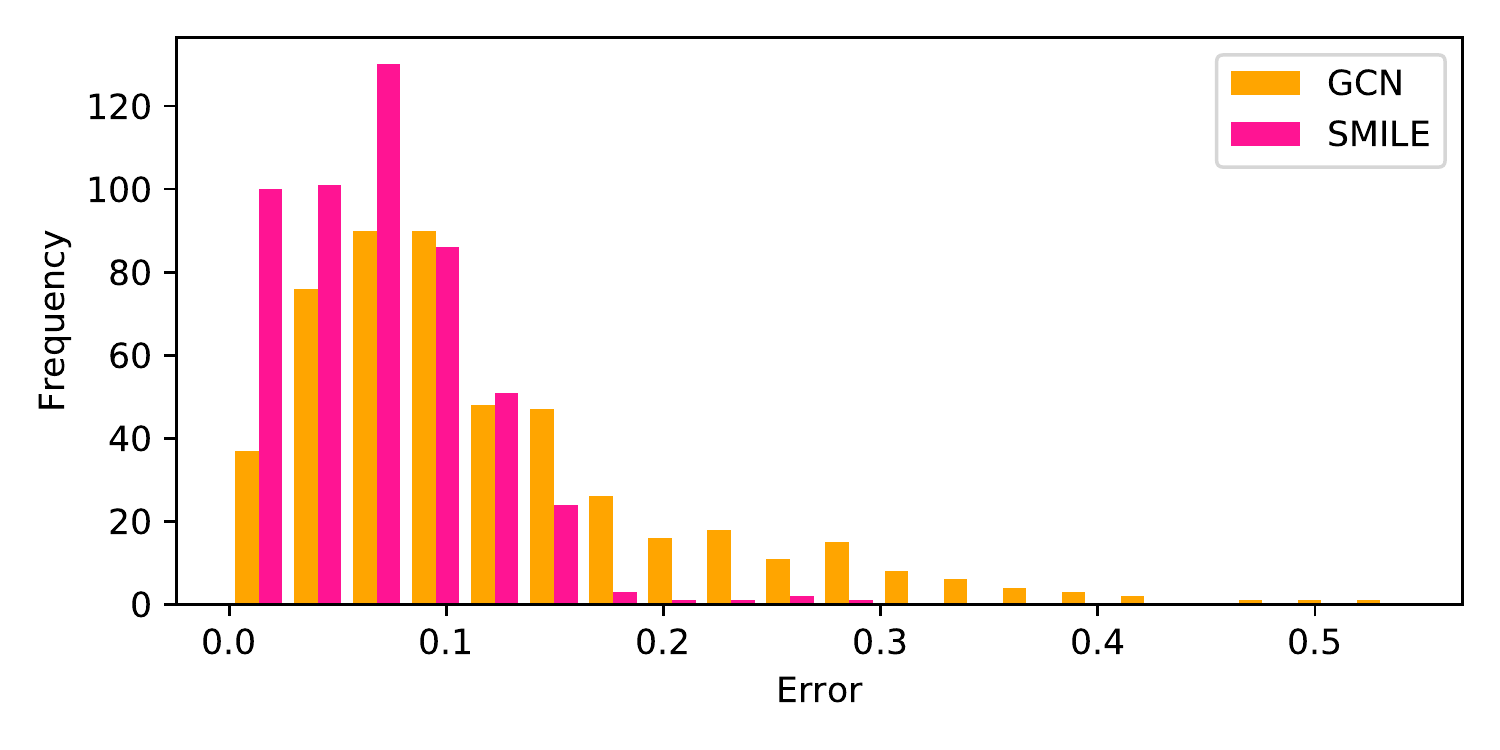}
    \caption{Error density for GCN and SMILE on the data from Fig. \ref{fig:compare_with_gcn}, where SMILE results in a more desirable distribution.}
    \label{fig:compare_density}
\end{figure}

\ph{Comparison with state of the art}
Firstly we consider the setting where $n=500, n_A=50, \sigma=0.1, p_\textrm{NLOS} = \frac{1}{10}$, and $\theta=d_\textrm{max}$, for which an example dataset is available\footnote{https://github.com/Yanzongzi/GNN-For-localization}. Fig. \ref{fig:compare_with_gcn} illustrates the performance of SMILE, which achieves an RMSE of 0.06 in 5.22 seconds, and GCN which achieves an RMSE of 0.11 in 5.73 seconds. GCN's performance is consistent with that reported in ~\cite{yan2021graph}, which reports RMSE for various other methods, including SDP with sparsity promoting regularization ~\cite{jin2020exploiting} which achieves an RMSE of 0.26. SMILE achieves the highest reported localization accuracy, and Fig. \ref{fig:compare_density} illustrates that not only is the error low on average but the error density has a smaller tail.

\begin{figure}
    \centering
    \includegraphics[width=\columnwidth]{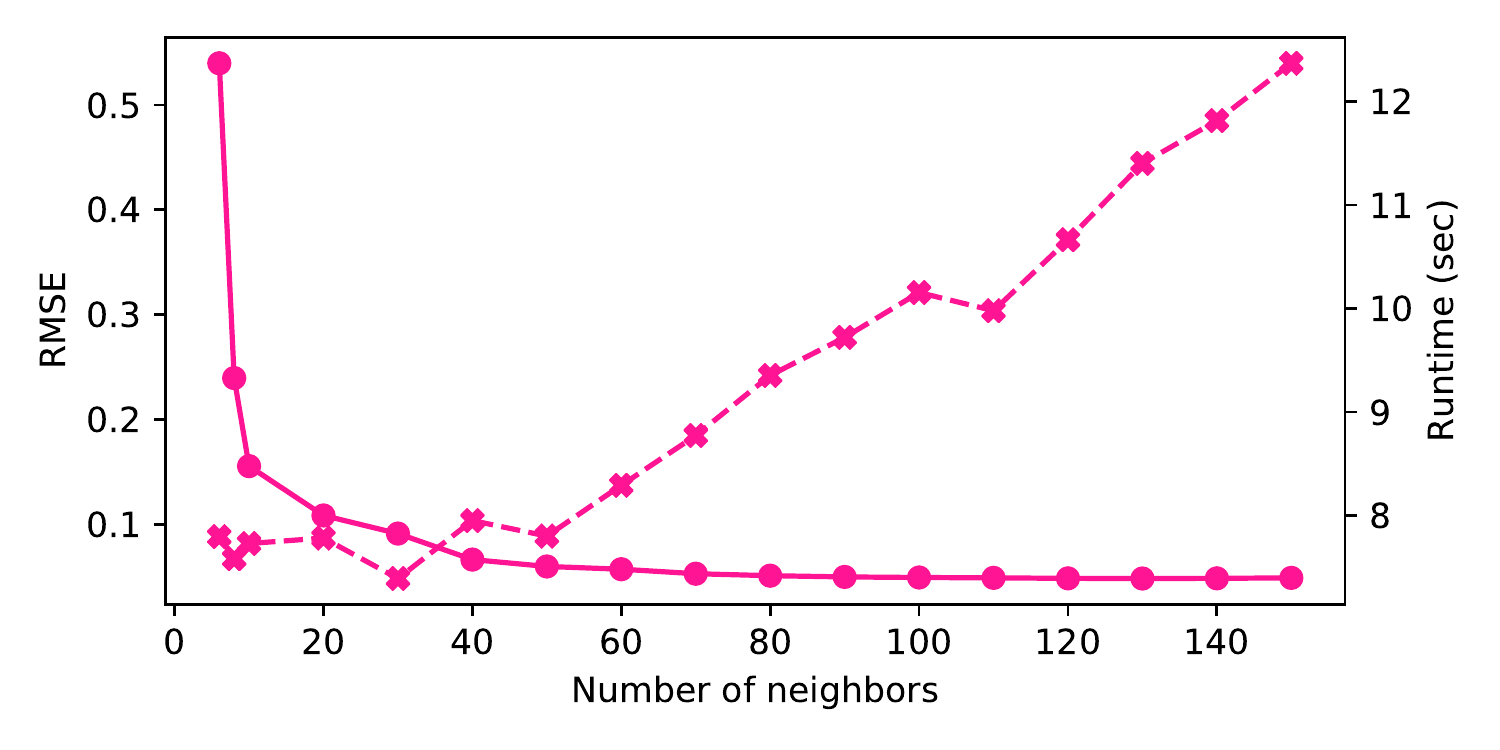}
    \caption{RMSE (solid line) and runtime (dashed line) trade-off as we vary the number of neighbors $k$. Each point is the average of 10 trials.}
    \label{fig:vary_n_neighbors}
\end{figure}

In Fig. \ref{fig:vary_n_neighbors}, we investigate the localization accuracy and computation time as we vary $k$, the number of neighbors used for linear embedding. We observe that RMSE remains consistent for $n_A \geq 50$. While the minimum of 0.048 is reached at $k=130$, this comes at the expense of increased computation time. Note that selecting $k$ is analogous to setting the GCN threshold, however we do not observe the degradation in performance that GCN is prone to when the threshold is too high. This comes from the aggregation component of convolution, which causes over-smoothing if a node has too many neighbors and results in the position estimates collapsing to a point. 

\begin{figure}
    \centering
    \includegraphics[width=\columnwidth]{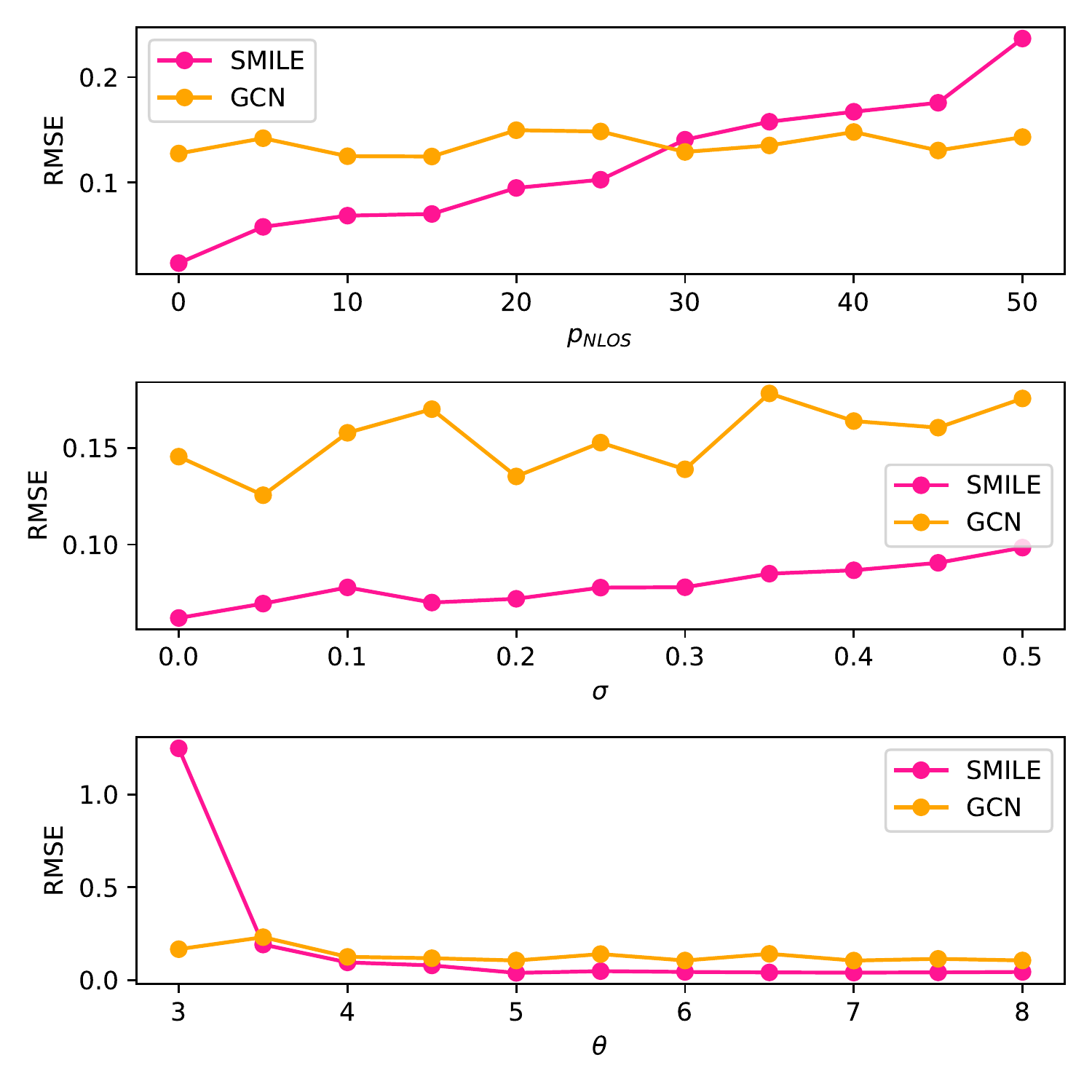}
    \caption{Performance of SMILE and GCN as the probability of NLOS links, standard deviation of Gaussian noise, and threshold for distances which can be measured are varied ($T=2$). }
    \label{fig:robustness_nlos}
\end{figure}

\ph{Robustness} 
Fig. \ref{fig:robustness_nlos} demonstrates the performance of SMILE and GCN as $p_\textrm{NLOS}$ varies from $0$ to $\frac{1}{2}$. From this, we observe that SMILE outperforms GCN for up to 30\% NLOS links. Beyond this, the assumption that $\mathbf{S}$ is sparse is no longer true, and performance suffers as expected. 
The same is true when $\theta$ falls below 3.5, and $(1-\mathbf{\Omega})$ is no longer sparse, i.e not enough entries in the observations matrix $\mbf{\tilde{O}}$.
Notably, we do not observe a clear trend in the performance of GCN as $\sigma$ increases. We posit that this is because the learned approach does not rely on exact decomposition, even in the absence of noise.

\begin{figure}
    \centering
    \includegraphics[width=\columnwidth]{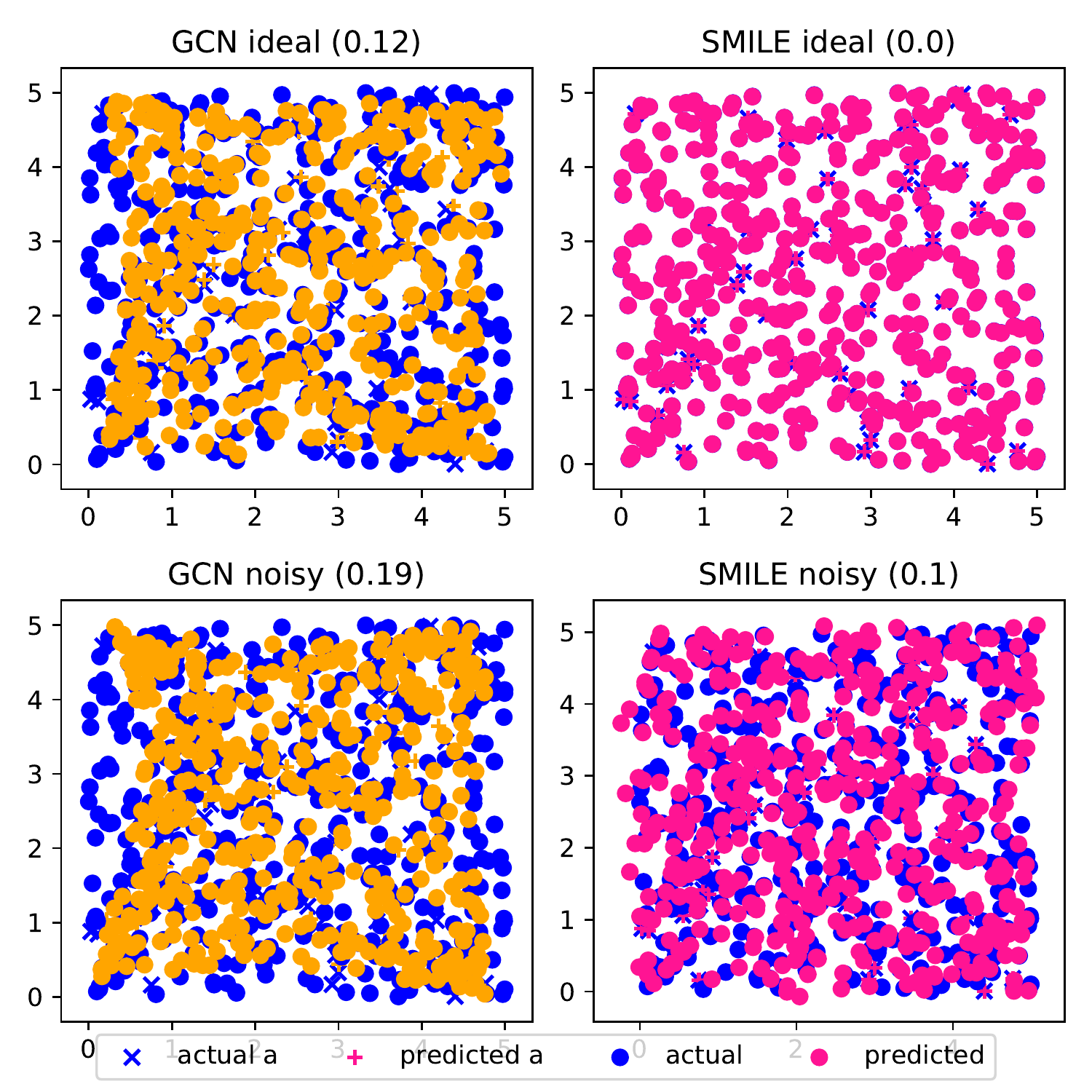}
    \caption{Performance of SMILE and GCN in an ideal setting ($p_\textrm{NLOS}=0, \sigma=0$) and a noisy setting ($p_\textrm{NLOS}=\frac{1}{10}, \sigma=0.5$). SMILE is robust to noise without sacrifice performance in ideal settings.}
    \label{fig:robustness}
\end{figure}

To test this, we compare the performance of GCN and SMILE on an ideal dataset and consider the results in Fig. \ref{fig:robustness}. While the performance of both methods is better in this ideal setting, we observe that GCN cannot exactly recover positions.
While being robust to Gaussian noise and sparse non-Gaussian noise, SMILE is also accurate in ideal scenarios. 

\begin{figure}
    \centering
    \includegraphics[width=\columnwidth]{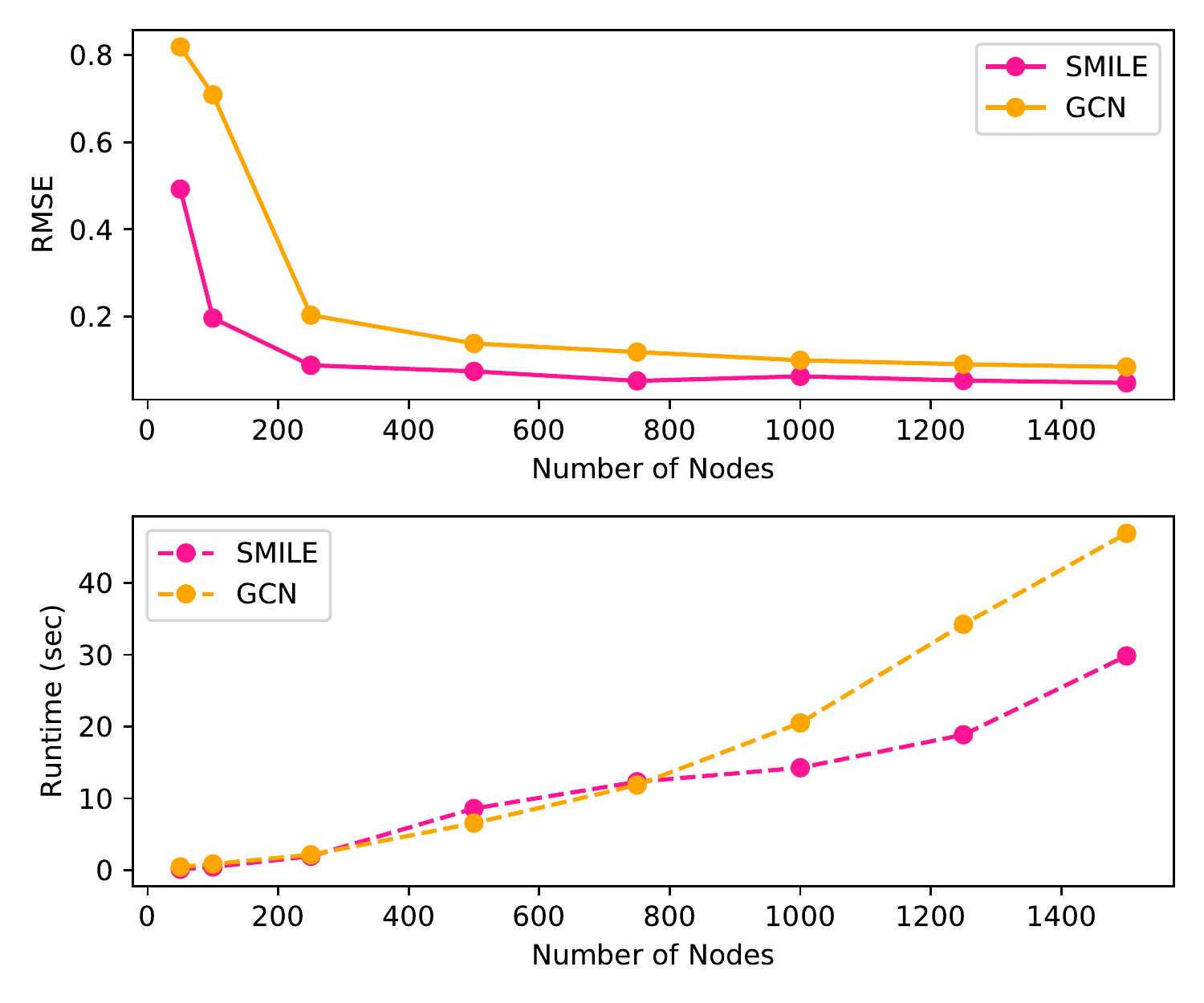}
    \caption{Compute time for different sizes of networks for GCN and SMILE. Previous work has shown that learned approaches (GCN, MLP, NTK) scale better than optimization approaches (LS, ECM, SDP) ~\cite{yan2021graph}. SMILE outperforms GCN for very large networks.}
    \label{fig:vary_num_nodes}
\end{figure}

\ph{Complexity}
As we increase the number of nodes, accuracy of both methods increases (accuracy similarly increases with the percentage of anchors). However, complexity increases with the size of the network. 
Because our approach is iterative, we measure complexity numerically (computation time) in lieu of analytically as it is difficult to predict when the sparse matrix inference will converge.
Fig. \ref{fig:vary_num_nodes} shows that the computation time of both SMILE and GCN remain reasonable as $n$ increases. 
For least squares optimization and SDP, we have seen that the compute time for very large networks becomes intractable ~\cite{yan2021graph}.
For the 1000 node network, the time to predict $\hat{\mathbf{P}}$ using SMILE is 29.84 seconds while the time to train and predict $\hat{\mathbf{P}}$ using GCN is 46.90 seconds. Note that this is a fair comparison because the learned approach requires training a specific model for each network; the same learned weights are not applicable to a new network, and neither is the SMILE weight matrix $\mathbf{H}$. 
The low compute times for SMILE are likely because the linear embedding component depends on the number of neighbors (rather than the number of nodes) and the sparse matrix inference component adapts the step size to the number of nodes.
We highlight that scalability is not at the cost of localization accuracy. For $n=1500$, SMILE achieves RMSE of 0.05 while GCN achieves RMSE of 0.08.


\begin{figure*}
    \centering
    \includegraphics[width=0.8\textwidth]{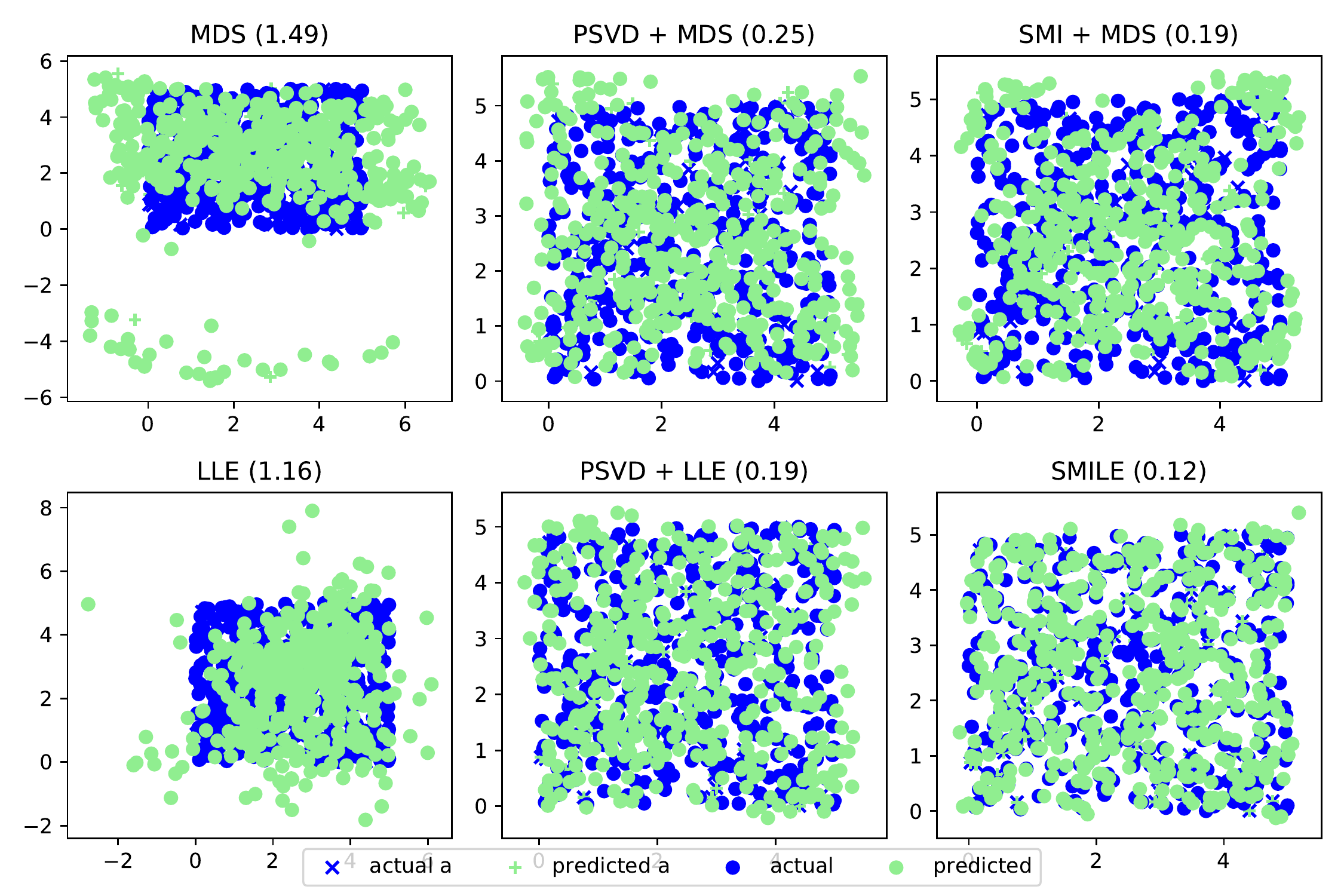}
    \caption{Ablation study with $\sigma=0.3, \theta=5$. Top row, left to right: classical MDS achieves RMSE 1.49, de-noising via PSVD improves RMSE to 0.25, and sparse matrix inference further improves RMSE to 0.19. Bottom row, left to right: LLE with direct position recovery achieves RMSE 1.16, de-noising via PSVD improves RMSE to 0.19, and and sparse matrix inference further improves RMSE to 0.12.}
    \label{fig:ablation}
\end{figure*}

\ph{Ablation}
Thus far we have compared our approach to an existing solution, but it is also interesting to consider the role of each component of SMILE and, in particular, whether simpler existing approaches from the literature are sufficient.
Fig. \ref{fig:ablation} considers a dataset with $\sigma = 0.3, p_\textrm{NLOS}=\frac{1}{10}$, and $\theta=5$.
We compare the performance of (A) no rank reduction, (B) rank reduction via PSVD, and (C) sparse and low-rank matrix recovery. In conjunction, we consider (1) classical MDS with the Kabsch algorithm for coordinate system registration ~\cite{bachrach2005localization} and (2) LLE-based embedding using anchor positions. 
Specifically, the top left plot represents a naive baseline: multidimensional scaling assuming Gaussian zero-mean noise. The bottom right plot represents SMILE.

We observe several key takeaways from this figure.
\begin{enumerate}
    \item \ph{De-noising} Moving from left to right, each column of this plot contains increasing more sophisticated noise reduction and decreases the final RMSE. In particular, sparse matrix inference increases localization accuracy by almost an order of magnitude.
    \item \ph{Nearest neighbors} Methods in the top row use all available links in eigen-decomposition, while methods in the bottom row use only the nearest neighbors to determine weights for linear reconstruction. This local-neighborhood approach consistently improves localization accuracy. Note that if $\theta \geq d_\textrm{max}$, i.e. no measurements are missing, the performance of (1) and (2) are similar. This is likely because LLE's strength comes from relying more on nearby measurements, and our approach to matrix completion puts missing measurements at effectively long distances. In setting where long-distance measurements are unreliable, which is commonly the case in realistic wireless communication ~\cite{schwengler2016wireless}, using nearest neighbors is advantageous.
    \item \ph{SMILE} The combination of both sparse matrix inference and linear embedding is more robust to NLOS attenuation and missing measurements than either component in isolation.
\end{enumerate}


\subsection{Experimental results}
In this section we apply SMILE to two real-world small scale datasets, and discuss its performance, our findings, and directions for future work.

\begin{figure}
    \centering
    \includegraphics[width=\columnwidth]{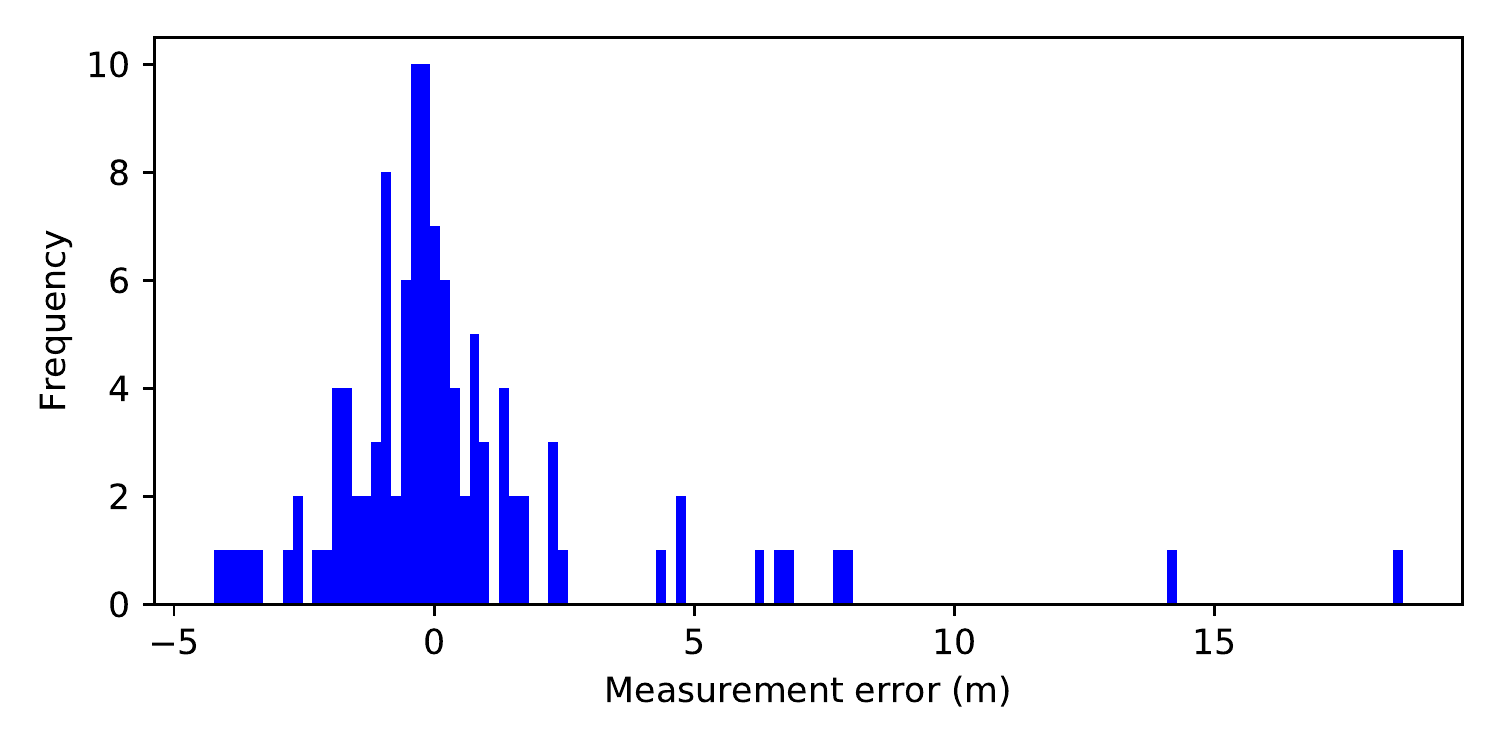}
    \caption{Measurement error for 11 node wireless sensor network outdoors, comparing distance from RSS (Eq. \ref{eq:RSS}) with ground truth.}
    \label{fig:cell_phone_error}
\end{figure}

\ph{Sensor network}
First we consider a network of 11 Mica2 motes placed randomly in a parking lot ~\cite{yedavalli2005ecolocation} \footnote{http://anrg.usc.edu/www/download\_files/RSSLocalizationDataSet\_11nodes.txt}. The maximum distance between any two nodes is 10.57m. We consider received signal strength (RSS) measurements between pairs (RSS is averaged over 20 packets). 
To estimate distance, we use the following log-distance path loss model
\begin{equation}
\label{eq:RSS}
    RSS(d) = P_\textrm{tx} - PL(d_0) - 10 \gamma  \log_{10}(\frac{d}{d_0}) + \mathcal{N}(0,\sigma^2)
\end{equation}
where $P_\textrm{tx}$ is the transmit power, $PL(d_0)$ is the path loss at a reference distance, and $\gamma$ is the path loss exponent ~\cite{schwengler2016wireless}.
For this data, use the model $\gamma = 2.9$ and $P_\textrm{tx} - PL(d_0) = -49.12 dB$ for $d_0=1$.
Given the ground truth location estimates, Fig. \ref{fig:cell_phone_error} shows that the noise in this dataset has mean 0.35m with standard deviation 2.94.
Note that even though we believe this data to be in open space (all LOS), seven links see a measurement error of greater than 5m.

\begin{figure}
    \centering
    \includegraphics[width=\columnwidth]{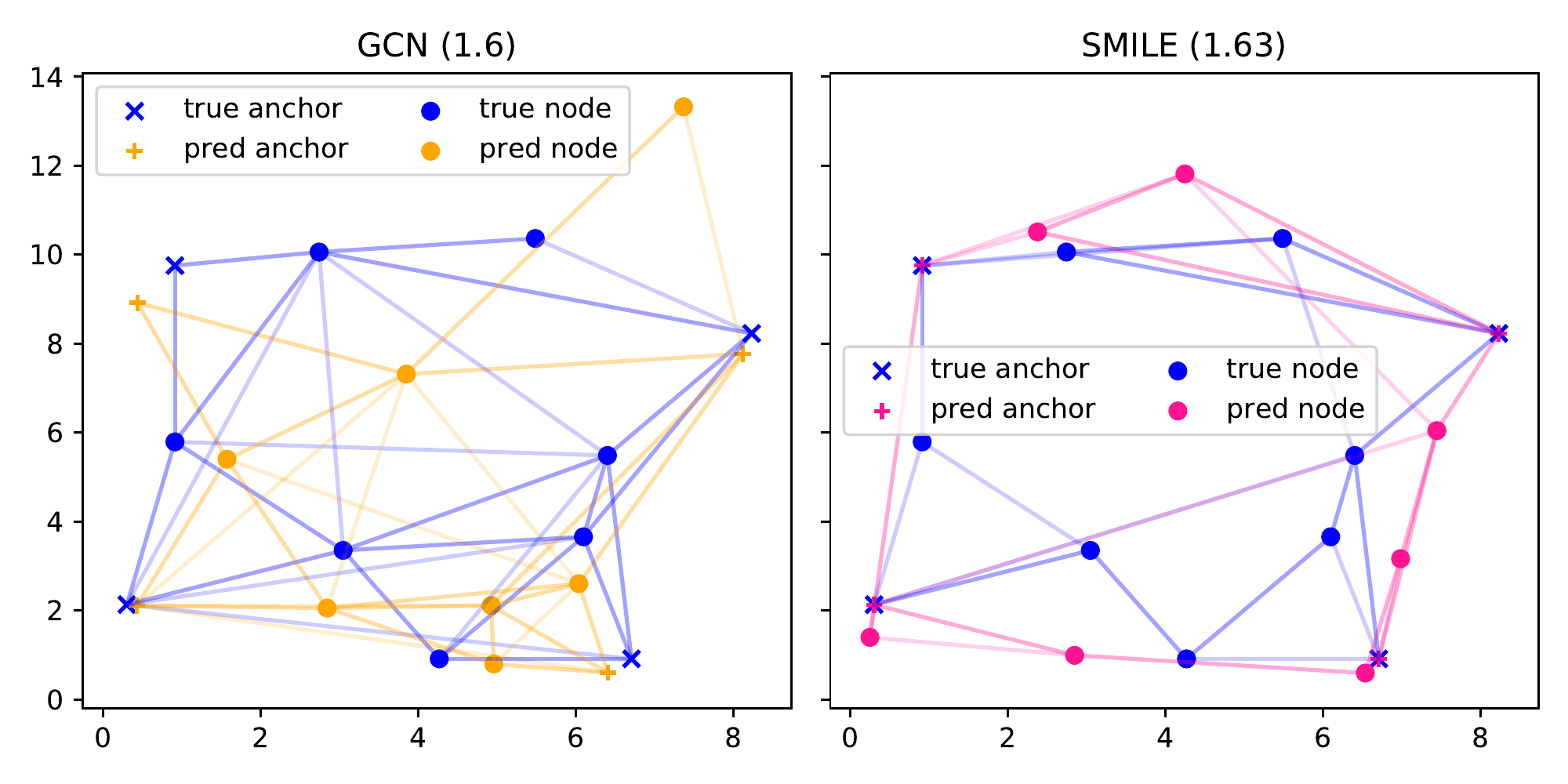}
    \caption{Localization accuracy of GCN (RMSE of 1.6) and SMILE (RMSE of 1.63) on the sensor network. Edges in the left graph represent observed distances less than $\theta_\textrm{GCn}$. Edges in the right graph connect each node $i$ to its nearest neighbors $NN(i)$.}
    \label{fig:cellphone}
\end{figure}

Figure \ref{fig:cellphone} shows the performance of GCN and SMILE on this network, where we update the SMILE parameters to  $\lambda=0.05$, $\mu=0.05$, and $T=10$ because the network is smaller.
If we assume four of these nodes are anchors, GCN achieves RMSE of 1.67 with threshold 5.2m, and SMILE achieves a comparable RMSE of 1.63 with $k=3$, guessing $\beta=11$. These results are the average performance of 10 trials, and the best parameters for each algorithm were selected empirically.

\begin{figure}
    \centering
    \includegraphics[width=\columnwidth]{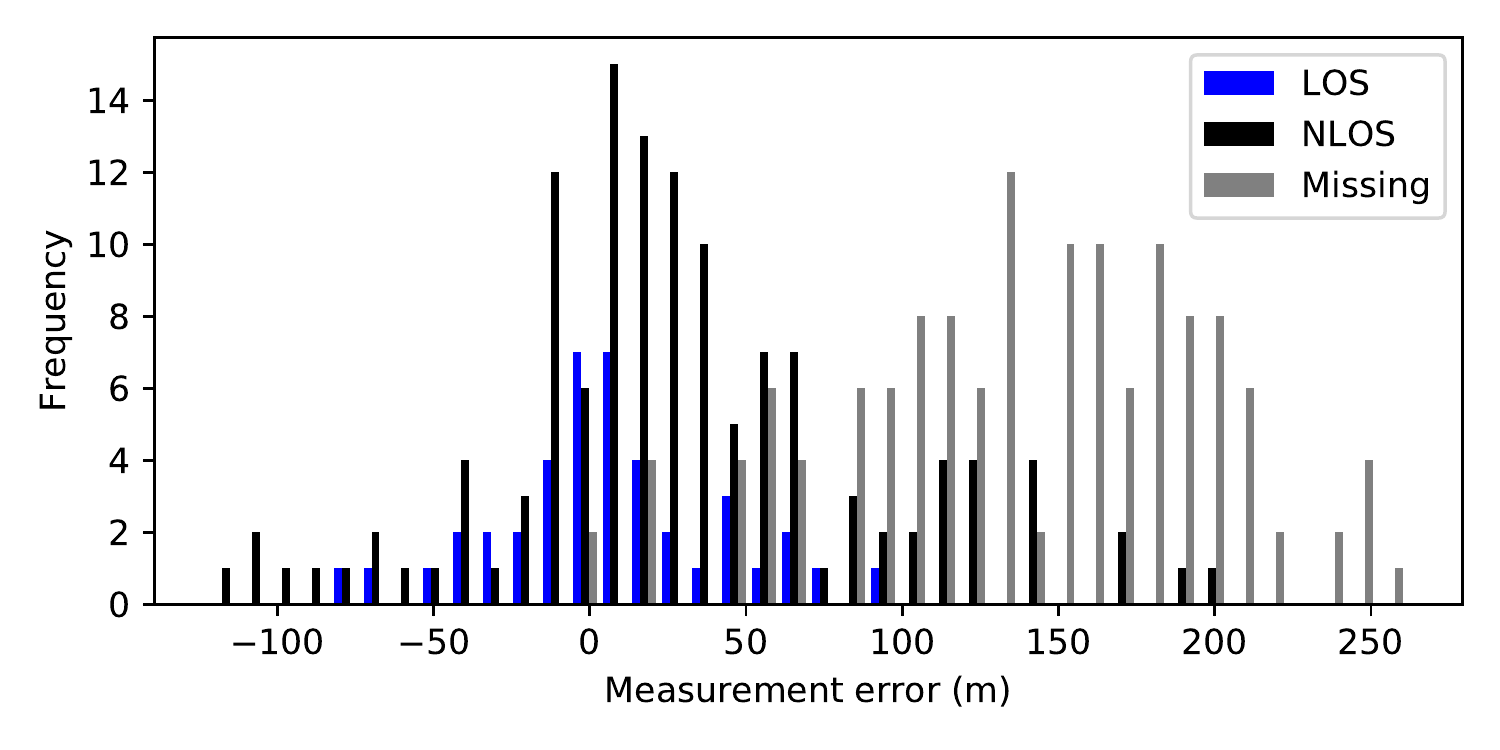}
    \caption{Measurement error for 18 node robotic network in GPS-denied environment.}
    \label{fig:robot_error}
\end{figure}

\ph{Robotic network} 
Secondly, we consider measurements from a robotic network with 18 nodes, 13 of which are stationary beacons at known locations, and 5 of which are mobile (legged and wheeled) robots\footnote{https://github.com/NeBula-Autonomy}. Each robot carries a Streamcaster 4200 from Silvus Technologies, which are also used as beacons. We consider a set of pairwise RSS measurements from a single timestamp of robotic exploration. They nodes are spread over a large area in an underground limestone mine with distances between nodes of up to 341 meters. 
Ground truth positions are available via a simultaneous localization and mapping algorithm, and we assume these are accurate ~\cite{chang2022lamp}.
Estimates of whether links are LOS is also available via a LiDAR-based predictive model ~\cite{clark2022prop}.
Fig. \ref{fig:robot_error} plots the error density for LOS, NLOS, and missing links from this data, where we note that unfortunately only 14\% of measurements are available and LOS. %
We augment this real dataset with simulated data such that missing measurements are sampled from a zero-mean Gaussian with standard deviation equivalent to the observed LOS noise ($\sigma = 36.13$).

\begin{figure}
    \centering
    \includegraphics[width=\columnwidth]{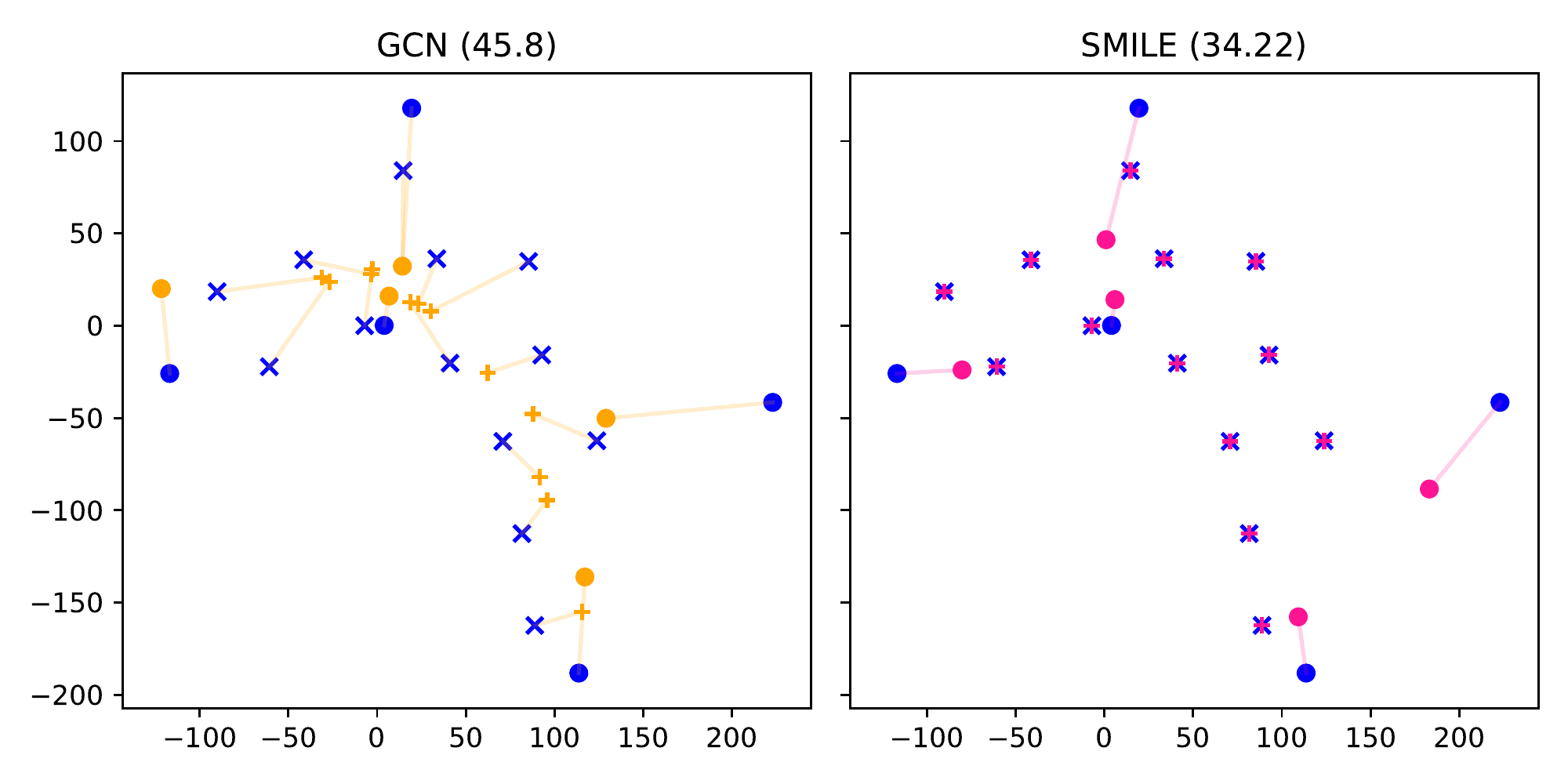}
    \caption{Localization accuracy of GCN (RMSE 45.8) and SMILE (RMSE 34.22) on the robotic network. Edges indicate error in distance.}
    \label{fig:robots}
\end{figure}

Fig. \ref{fig:robots} show the results of our approach on this realistic dataset, where edges indicate error in distance.
GCN achieves RMSE of 45.8 on this simulated data with threshold 110, while SMILE achieves an improved RMSE of 34.22 for $k=17$. These results are the average performance of 10 trials, and the best parameters for each algorithm were selected empirically.
We observe that the average localization error for unknown nodes is roughly the standard deviation of the noise on LOS links, while the standard deviation of NLOS error is 59.60 (mean 30.14m). This indicates that SMILE is able to achieve localization accuracy comparable to a distance-based model on a single LOS link for all nodes, even those which have many NLOS neighbors, and shows promising performance.

During robotic exploration, these anchors were deployed autonomously ~\cite{saboia2022achord, funabiki2020range}. 
This means anchors are only located in places the robots have already visited, while the exploration objective encourages the robots to move away from these anchors. In fact, the exploring robots then tend to be outside the convex hull defined by the anchors. This configuration appears to stress network localization, especially for smaller networks with significant noise. Prior work exists in robotic motion which minimizes localization uncertainty ~\cite{cano2022ranging, clark2021queue}, and network localization algorithms which specifically address this geometric setting may be an interesting direction for future work.

\subsection{Discussion}

One advantage of SMILE is that we compute $\mathbf{Y}$ which approximates the sparsity structure of $\mathbf{S}$. Therefore, we can use this method to estimate which links are NLOS and where walls, obstacles, or other potentially adversarial sources of attenuation may be present. This means SMILE could be useful as a complementary modality for simultaneous localisation and mapping (SLAM) algorithm for multi-robot systems, and has potential applications in robotic exploration for disaster mitigation and military applications ~\cite{clark2022prop}.

While it is impossible to unique localize a network in the absence of anchors (given the possibility of translations and rotations), anchor-less localization problems have significant overlap with other useful signal processing problems. In face, SMILE could be directly applied to anchor-less localization problems by replacing lines 29-30 in the algorithm with the embedding step of conventional LLE, as LLE typically does not require anchors ~\citep{saul2000introduction}. Thus, our approach may be applicable to problems outside the sensor networks domain. 
For instance, consider a matrix of user data, where each row corresponds to a specific user and a small number of users obfuscate their data with the addition of non-negative noise. SMILE could be applied to the problem of determining how similar users are, by extracting the sparse obfuscation matrix and finding an embedding of users in low-dimensional space.

We considered a modification to LLE in which we replace the $k$ nearest neighbors with the $n_A$ anchors as reference points. We expected this would minimize the risk of errors propagating, as each of the anchors is at a known location. Interestingly, we did not see an improvement in performance.
Similarly, we considered a modification to the sparse matrix inference approach in which we project $\mathbf{X}$ and $\mathbf{Y}$ on to the space of matrices which are non-negative and symmetric. We expected this would improve performance as $\mathbf{D}$ and $\mathbf{S}$ are non-negative and symmetric. Preliminary experiments did not demonstrate a significant improvement in localization accuracy, and in some cases accuracy suffered.
Low-rank matrix approximation which constrains the solution to have the characteristics of Euclidean distance matrix, without sacrificing localization accuracy, is a direction for further study.


Following ~\cite{yan2021graph}, we considered the probability of each link being NLOS as independent. However, in realistic settings there will be some correlation based on the structure of the environment itself. For example, in Fig. \ref{fig:illustration} we illustrate 20 nodes in an indoor setting with three rooms and one hallway. Nodes in the smaller rooms are more subject to NLOS attenuation, given their proximity to two walls. Currently, we estimate that links with the weakest RSS are most likely to be NLOS. However it may instead be the case that certain nodes are more likely to have NLOS links, is in Fig. \ref{fig:illustration}. Exploiting patterns like this in the structure of $\mathbf{S}$ is an interesting direction for future work.


\section{Conclusion}
\label{sec:conclusion}
We present Sparse Matrix Inference and Linear embedding, a novel network localization algorithm which is robust to noisy, NLOS, and missing measurements.
Our approach outperforms the state of the art on large, simulated networks, achieving high localization accuracy with low computation times.
We see promising results for small networks in real-world settings, and in the future we'd like to collect real-world data for larger networks. Other directions for future work include extensions to make the approach distributed~\cite{Biswas2006approach,costa2006distributed}, or consider time-series RSSI measurements and Bayesian estimation ~\cite{wymeersch2009cooperative,jin2020bayesian}.

\section*{Acknowledgments}
Thanks to Kiran Yedavalli for the sensor network data and Team CoSTAR for the robotic network data. This work was funded in part by NASA Space Technology Research Fellowship Grant No. 80NSSC19K1189.

\bibliographystyle{ACM-Reference-Format}
\bibliography{ref}

\end{document}